\documentclass[%
reprint,
 amsmath,amssymb,
 aps,
prd,
]{revtex4-1}

\usepackage{graphicx}
\usepackage{dcolumn}
\usepackage{bm}
\usepackage{color}

\def\apj{Astrophys. J.~}
\def\apjs{Astrophys. J. Suppl.~}
\def\apjl{Astrophys. J. Lett.~}
\def\mnras{Mon. Not. R. Astr. Soc.~}
\def\prd{Phys. Rev. D~}

\def\physrep{Phys. Rep.~}
\def\araa{Ann. Rev. Astr. \& Astrophys.~}
\def\aap{Astr.\& Astrophys.~}

\def\jcap{J. Cosmo. \& Astroparticle Phys.~}
\def\nat{Nature}

\def\om{\Omega_{\rm M}}
\def\olam{\Omega_{\Lambda}}
\def\kms{\rm km\,s^{-1}}
\def\ergs{\rm erg\,s^{-1}}

\def\enu{E_{\nu}}

\def\ep{E_{\rm p}}

\def\tacc{t_{\rm acc}}
\def\tpp{t_{pp}}
\def\tpg{t_{p\gamma}}
\def\tdyn{t_{\rm dyn}}

\def\fkin{f_{\rm kin}}
\def\ksib{\xi_{\rm B}}

\def\lbol{L_{\rm bol}}
\def\dl{D_{\rm L}}
\def\ep{E_{\rm p}}
\def\eph{E_{\rm ph}}
\def\gp{\Gamma_{\rm p}}
\def\np{n_{\rm p}}
\def\ns{n_{\rm s}}
\def\ts{T_{\rm s}}
\def\Nph{N_{\rm ph}}
\def\sigpp{\sigma_{pp}}
\def\kpp{\kappa_{pp}}
\def\epk{\bar{\epsilon}_{\rm pk}}
\def\sigpk{\sigma_{\rm pk}}
\def\kpk{\kappa_{\rm pk}}
\def\eg{E_{\gamma}}

\def\lbol{L_{\rm bol}}
\def\tsal{t_{\rm sal}}
\def\ent{\epsilon_{\rm nt}}

\def\vs{v_{\rm s}}
\def\rs{R_{\rm s}}
\def\emax{E_{\rm max}}

\def\zref{z_{\rm ref}}
\def\msun{M_{\odot}}
\def\lsun{L_{\odot}}

\begin{document}

\preprint{APS/123-QED}

\title{Cumulative neutrino background from quasar-driven outflows}

\author{Xiawei Wang}
\email{xiawei.wang@cfa.harvard.edu}
\author{Abraham Loeb}%
\affiliation{Department of Astronomy, Harvard University, 60 Garden Street, Cambridge, MA 02138, USA}%

             
\begin{abstract}
Quasar-driven outflows naturally account for the missing component of the extragalactic $\gamma$-ray background through neutral pion production in interactions between protons accelerated by the forward outflow shock and interstellar protons. We study the simultaneous neutrino emission by the same protons.
We adopt outflow parameters that best fit the extragalactic $\gamma$-ray background data and derive a cumulative neutrino background of $\sim10^{-7}\,\rm GeV\,cm^{-2}\,s^{-1}\,sr^{-1}$ at neutrino energies $\enu\gtrsim 10$ TeV, which naturally explains the most recent IceCube data without tuning any free parameters.
The link between the $\gamma$-ray and neutrino emission from quasar outflows can be used to constrain the high-energy physics of strong shocks at cosmological distances.
\end{abstract}
\pacs{Valid PACS appear here}
\maketitle
%
\textit{Introduction.}---There is currently strong observational evidence for the existence of large-scale outflows driven by the active galactic nuclei (AGN), including the presence of broad absorption lines in quasars \cite{zakamska2014, arav2015} and multiphase outflows in nearby ultraluminous infrared galaxies (ULIRGs) \cite{rupke2011, tombesi2015}.
Semi-relativistic winds with a speed of $\sim 0.1\,c$ are typically produced by quasars in the surrounding interstellar medium, driving a forward shock that accelerates a swept-up shell accompanied by a reverse shock that decelerates the wind itself \cite{FGQ2012, king2015}.

In a previous paper, we derived a detailed hydrodynamical model for the quasar outflow's interaction with the ambient medium \cite{wang2015} (see Supplemental Material), including a disk and a halo components for the host galaxy gas. 
The gas density profile was self-consistently determined by the halo mass and redshift.
The continuous energy injection was assumed to be a fraction of the quasar's bolometric luminosity $\fkin\lbol$ during the quasar's lifetime, which is of order the Salpeter time $\tsal\sim 4\times 10^7$ yrs for a radiative efficiency of 0.1 \cite{yu2002}.
In the upper panel of Fig.2, we show the forward shock velocity $\vs$ as a function of radius $\rs$ for the outflow in a dark matter halo of mass $\sim 10^{12}\,\msun$ at a redshift of $z\sim 0.1$. 
We find that $\vs\gtrsim 10^{3}\,\kms$ within the galactic disk with a decline to few hundreds $\kms$ when the outflow reaches the edge of the halo.
In analogy with supernova (SN) remnants \cite{caprioli2012, ackermann2013}, protons should be accelerated via Fermi acceleration to relativistic energies in the forward outflow shock.

The resulting proton number density per unit volume per unit energy can be expressed as a power-law with an exponential high-energy cutoff:
\begin{equation}
\frac{dN_{\rm p}}{d\ep}=N_0\ep^{-\gp}\,\exp\left(-\frac{\ep}{\emax}\right)\;,
\end{equation}
where $\gp$ is the power-law index, $N_0$ is the normalization constant and $\emax$ is the maximum energy of the accelerated protons.
The value of $N_0$ can be obtained by setting $\int N_{\rm p}(\ep)\ep d\ep=\frac{3}{2}\ent\ns k\ts$, where $\ent$ is the fraction of energy that goes to accelerated protons and $\ns$ and $\ts$ are the number density and temperature of the shocked medium, respectively.
$\emax$ can be obtained by equaling the acceleration time scale, $\tacc$, and the minimum between the cooling timescale and the dynamical timescale, $\tdyn\sim\rs/\vs\approx10^6\,R_{\rm s,kpc}v_{\rm s,3}^{-1}$ yrs.
We adopt $\tacc\sim\ep c/eB\vs^2\approx300\,E_{\rm p,TeV}B_{-6}^{-1}v_{\rm s,3}^{-1}$ yrs, where $B$ is the post-shock magnetic field \cite{blandford1987}. 
Here $E_{\rm p,TeV}=(\ep/\rm TeV)$, $v_{\rm s,3}=(\vs/10^3\rm \kms)$, $R_{\rm s,kpc}=(\rs/\rm kpc$) and $B_{-6}=(B/\mu\rm G)$.
We assume that a fraction of the post shock thermal energy is carried by the magnetic field, giving $B=(8\pi\ksib\ns k\ts)^{1/2}$, with a value $\ksib=0.1$ calibrated based on SN remnants \cite{chevalier1998}.
Protons may lose energy via synchrotron, inverse Compton scattering, hadro-nuclear ($pp$) or photo-hadronic ($p\gamma$) processes.
As discussed later, $pp$ collisions provide the dominant cooling mechanism for protons.
The corresponding timescale, $\tpp$, can be written as \cite{kelner2006}:
\begin{equation}
\tpp^{-1}=\np\sigpp c \kpp\;,
\end{equation}
where $\kpp\sim0.5$ is the inelasticity parameter and $\sigpp$ is the cross section for $pp$ collisions \cite{kelner2006}:
\begin{equation}
\sigpp = (34.3+1.88\ell+0.25\ell^2)\left[1-\left(\frac{E_{\rm th}}{\ep}\right)^4\right]^2\,\rm mb\;,
\end{equation}
with $\ell=\ln E_{\rm p,TeV}$ and $E_{\rm th}\approx1.22$ GeV being the threshold energy for $pp$ collisions.
For $B_{-6}=1$, $v_{\rm s,3}=1$ and $R_{\rm s,kpc}=1$, we find $\emax\sim10^6$ GeV.
The parameters $\fkin$ and $\ent$ and $\gp$ constrain the hadronic emission from quasar outflows.

$\gamma$-ray emission is produced via the decay of the neutral pions generated in $pp$ collisions, $\pi^0\rightarrow2\gamma$.
The detailed calculation of the integrated $\gamma$-ray background is discussed in our companion paper \cite{wang2016}.
While blazars account for $\sim 50\%$ of the extragalactic $\gamma$-ray background (EGB) at $\eg\lesssim 10$ GeV and almost all the EGB at higher energies \cite{ajello2015}, we use parameter values consistent with outflow observations \cite{tombesi2015} and find that our model produces $\gamma$-ray emission that make up $\sim 30\%$ of the EGB at $\eg\lesssim 10$ GeV and matches the required spectral shape of the EGB. 
The $\gamma$-ray emission by quasar outflows dominates over radio galaxies and star-forming galaxies, based on the most recent \textit{Fermi}-LAT data \cite{ackermann2015} and previous studies \cite{inoue2011, ackermann2012, dimauro2014, wang2016}.
For a given $\Gamma_{\rm p}$, we can fix the free parameters in our model, $\fkin\ent$, by fitting the EGB data.
For $\ent\sim10\%$, we find that $\fkin\sim3\%$, in agreement with observations of outflows \cite{tombesi2015}.
%
%
\\
\\
\indent
\textit{Neutrino production.}---Next, we calculate the simultaneous neutrino emission from the same protons, which lose energy via two main channels of pion production: $p+\gamma\rightarrow p+\pi^0$ or $n+\pi^+$ and $p+p\rightarrow \pi^+ + \pi^- + \pi^0$.
In the $p\gamma$ channel, relativistic protons lose energy by interacting with X-ray photons from the hot coronae above the accretion disk.
The timescale for $p\gamma$ interactions is given by \cite{stecker1968, murase2012}:
\begin{equation}
\begin{split}
&\tpg^{-1}=\frac{c}{2\gamma_{\rm p}^2}\epk\Delta\epk\sigpk\kpk
\\
&\times\int_{\epk/2\gamma_{\rm p}}^{\infty}\,\frac{(d\Nph/d\eph)}{\eph^2}\,d\eph\;,
\end{split}
\end{equation}
where $\epk\sim0.3$ GeV, $\sigpk\sim5\times10^{-28}\,\rm cm^2$, $\kpk\sim0.2$, $\Delta\epk\sim0.2$ GeV, $\gamma_{\rm p}=\ep/m_{\rm p} c^2$, and $d\Nph/d\eph$ is the number density of soft photons per photon energy.
Assuming an X-ray luminosity $L_{\rm X}\sim 0.1\lbol$ and a power-law template with a spectral index of $\sim2.5$ for $d\Nph/d\eph$ \cite{elvis1994, marconi2004}, we estimate that $\tpg\sim10^{12}$ yrs $\gg\tpp\sim10^{8}$ yrs for $\ep\sim1$ PeV accelerated by a 10-kpc scale outflow from a quasar with a bolometric luminosity, $\lbol\sim10^{46}\,\ergs$. 
A detailed comparison of these timescales as a function of $\ep$ is shown in Fig.1, where we find that $\tpg$ is indeed substantially longer than $\tpp$ for $\ep\lesssim\emax$.
Therefore, we neglect $p\gamma$ interactions and consider $pp$ collisions as the dominant channel for proton cooling.
We have also verified that synchrotron and inverse Compton cooling of protons are negligible \cite{sturner1997}.
\begin{figure}
\includegraphics[angle=0,width=0.9\columnwidth]{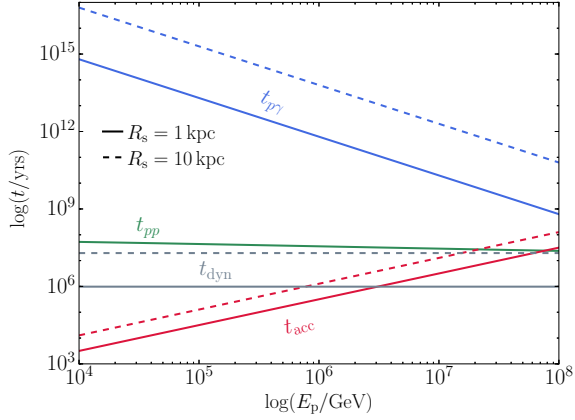}
\caption{Comparison of relevant timescales for the acceleration of protons, $\tacc$, the dynamics of the outflow shock, $\tdyn$ and for $pp$ and $p\gamma$ interactions, represented by the red, grey, green and blue lines, respectively.
The solid and dashed lines correspond to cases where the outflow propagates to distances of 1 kpc and 10 kpc, respectively.
We assume a quasar bolometric luminosity of $10^{46}\,\ergs$ and a magnetic field of 1 $\mu$G.
}
\end{figure}

Neutrinos are generated via the decay of charged pions, $\pi^+\rightarrow \mu^+ + \nu_{\mu} \rightarrow e^+ + \nu_e+\bar{\nu}_{\mu}+\nu_{\mu}$ and $\pi^-\rightarrow \mu^- + \bar{\nu}_{\mu} \rightarrow e^- + \bar{\nu}_e +\nu_{\mu}+\bar{\nu}_{\mu}$.
At the source, the production flavor ratio of neutrinos is $(\nu_e:\nu_{\mu}:\nu_{\tau})=(\bar{\nu}_e:\bar{\nu}_{\mu}:\bar{\nu}_{\tau})=(1:2:0)$, where $\nu_e$, $\nu_{\mu}$ and $\nu_{\tau}$ are electron, muon and tau neutrinos, respectively.
Neutrino oscillations on the way to Earth results in equal numbers of $\nu_e$, $\nu_{\mu}$ and $\nu_{\tau}$.
We consider $\nu$ and $\bar{\nu}$ equally since terrestrial neutrino detectors do not distinguish between them \cite{crocker2000}.

The neutrino spectrum from an individual outflow is given by:
\begin{equation}
\begin{split}
\Phi_{\nu}(\enu)=c\np\int_0^{1}&\sigpp(\enu/x)\frac{dN_{\rm p}}{d\ep}(\enu/x)\\
&F_{\nu}(x,\enu/x)\frac{dx}{x}\;,
\end{split}
\end{equation}
where $x=\enu/\ep$ and $F_{\nu}$ is the neutrino spectrum calculated based on the prescription given by Ref. \cite{kelner2006} (see Supplemental Material for details).
In the lower panel of Fig.2, we show the resulting neutrino spectrum when an outflow propagates to the edge of the galactic disk $R_{\rm d}$.
The flux is sensitive to $\gp$ in that a steeper proton spectrum leads to fewer neutrinos with energies above 1 TeV.
Note that the neutrino flux drops significantly as the outflow propagates outside the galactic disk, due to the declines in the shock velocity and the ambient gas density.
\begin{figure}
\includegraphics[angle=0,width=0.8\columnwidth]{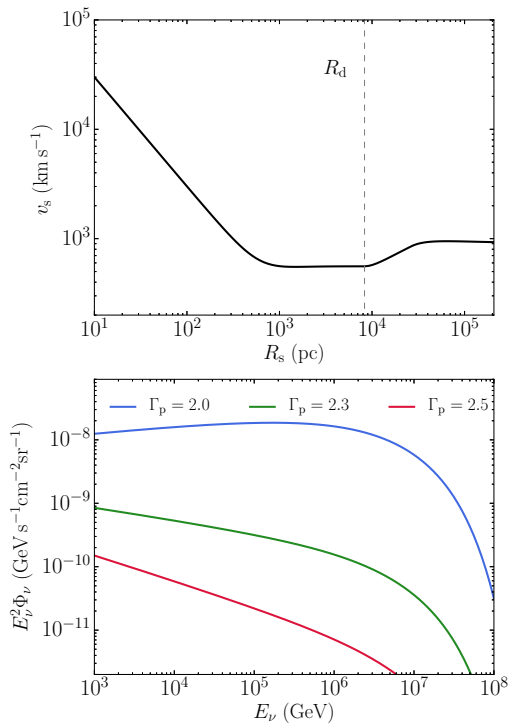}
\caption{Quasar outflow speed vs distance and corresponding neutrino flux summed over all flavors from $pp$ collisions.
The outflow is hosted by a $10^{12}\msun$ halo at redshift of $z=0.1$.
In the upper panel, we show the speed of the outflowing shell, $\vs$, as a function of its radius, $\rs$.
The dashed vertical line marks the location of the galactic disk, $R_{\rm d}$.
The lower panel shows the neutrino flux from $pp$ collisions as the outflow propagates to the edge of the disk.
The blue, green and red lines correspond to different values of the power-law index of the accelerated protons, namely $\gp=2.0$, 2.3 and 2.5, respectively.
}
\end{figure}
%
%
\\
\\
\indent
\textit{Cumulative neutrino background.}---The integrated neutrino flux from quasar outflows can be obtained by summing the neutrino emission over the entire quasar population at all bolometric luminosities, $\lbol$, and redshifts, $z$,
\begin{equation}
\begin{split}
\enu^2\Phi_{\nu}=\iint&\Phi(\lbol,z)\frac{\bar{L}_{\nu}(\enu^\prime,\lbol,z)}{4\pi\dl^2(z)}\\
&\times d\log\lbol\frac{dV}{dzd\Omega}dz\,
\end{split}
\end{equation}
where $V$ is the comoving cosmological volume, $E^\prime_{\nu}=\enu(1+z)$ is the neutrino energy at the source frame, and $\bar{L}_{\nu}=\tsal^{-1}\int L_{\nu}(\enu,\lbol,z,t)\,dt$ is the time-averaged neutrino flux from an individual source.
$\Phi(\lbol,z)$ is the bolometric luminosity function, given by \cite{hopkins2007}:
\begin{equation}
\Phi(\lbol,z)=\frac{\Phi_{\star}}{(\lbol/L_{\star})^{\gamma_1}+(\lbol/L_{\star})^{\gamma_2}}\;,
\end{equation}
where $L_{\star}$ varies with redshift according to the functional dependence,  
$\log L_{\star}=(\log L_{\star})_0+k_{L,1}\xi+k_{L,2}\xi^2+k_{L,3}\xi^3$, $\xi=\log[(1+z)/(1+\zref)]$, with $\zref=2$ and $k_{L,1}$, $k_{L,2}$ and $k_{L,3}$ being free parameters.
We adopt parameter values of the pure luminosity evolution model, where $\log(\Phi_{\star}/\rm Mpc^{-3})=-4.733$, $(\log (L_{\star}/\lsun))_0=12.965$, $\lsun=3.9\times10^{33}\,\ergs$, $k_{L,1}=0.749$, $k_{L,2}=-8.03$, $k_{L,3}=-4.40$, $\gamma_1=0.517$ and $\gamma_2=2.096$.
The comoving volume per unit solid angle can be expressed as:
\begin{equation}
\frac{dV}{dzd\Omega}=D_{\rm H}\frac{D_{\rm L}^2(z)}{(1+z)^2E(z)}\;,
\end{equation}
where $D_{\rm H}=c/H_0$ and $E(z)=\sqrt{\om(1+z)^3+\olam}$.
We adopt the standard cosmological parameters: $H_0=70\,\kms\rm Mpc^{-1}$, $\om=0.3$ and $\olam=0.7$ and integrate over the bolometric luminosity range of $\lbol=10^{42}-10^{48}\,\ergs$ and the redshift range of $z=0-5$.

Figure 3 shows the cumulative neutrino background (CNB) from quasar-driven outflows compared to the most recent IceCube data, which are fitted by two separate models \cite{aartsen2015}:
a differential model fitted by nine free parameters (indicated as the black points with error bars), and a single power-law model (indicated as the gray shaded region) in the form of $\Phi_{\nu}^{\rm pl}=\phi\times\left(\enu/100\,\rm TeV\right)^{-\gamma}$ where $\phi=6.7^{+1.1}_{-1.2}\times10^{-18}\,\rm GeV^{-1} cm^{-2} s^{-1} sr^{-1}$ and $\gamma=2.50\pm0.09$.
\begin{figure*}
\includegraphics[angle=0,width=1.9\columnwidth]{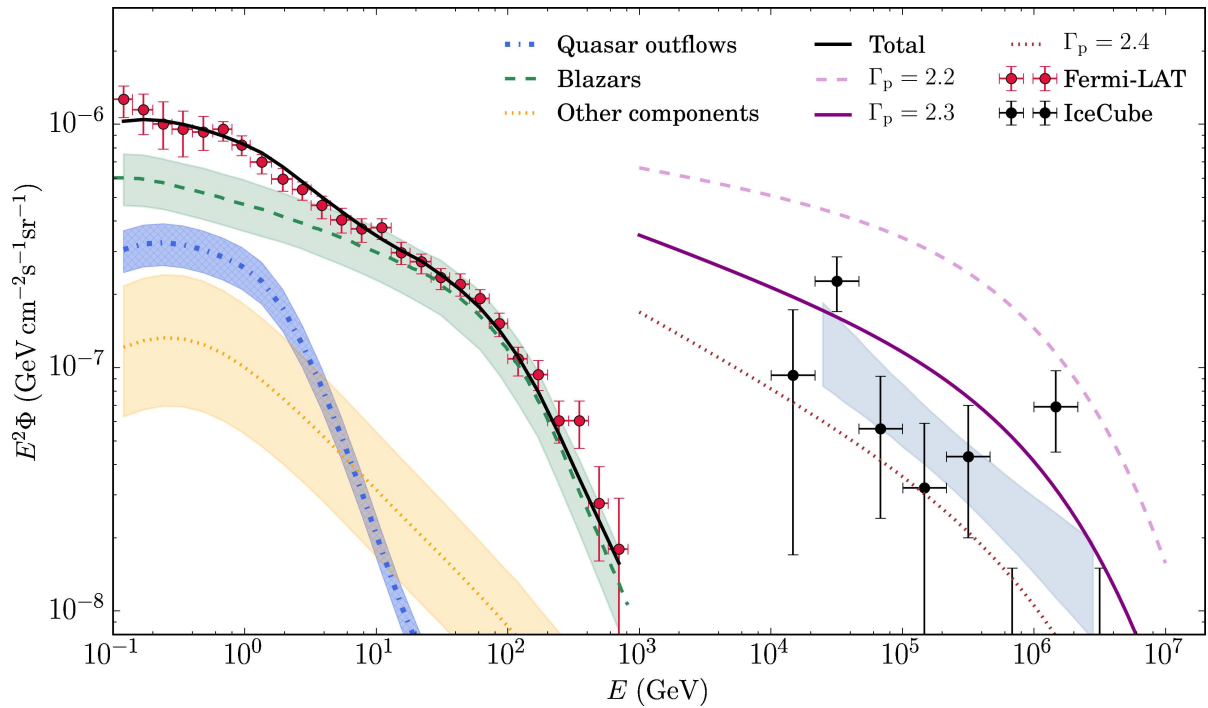}
\caption{Cumulative $\gamma$-ray (left) and neutrino background (right) from quasar-driven outflows.
The red points with error bars on the left are the observed data points for the $\gamma$-ray background from \textit{Fermi}-LAT \cite{ackermann2015}.
The blue, green and orange shaded regions correspond to the contribution from quasar outflows, blazars and other components (including radio galaxies and star-forming galaxies), respectively, and the total contribution from all components is represented by the solid black line.
The power-law and differential model of IceCube neutrino data (all flavors combined) are shown on the right as the gray shaded region and the black points with error bars, respectively \cite{aartsen2015}.
The pink, purple and brown lines correspond to the cumulative neutrino flux produced by quasar outflows where the accelerated protons have an energy distribution with a power-law index of $\gp=2.2$, 2.3 and 2.4, respectively.
}
\end{figure*}

For each value of $\gp$, we fix $\ent\fkin$ based on the best fit to the EGB and produce the neutrino background without allowing additional freedom in the parameter choices.
Interestingly, we find that the resulting flux explains the neutrino background observed by IceCube for $\gp\approx2.2-2.4$, which is the range of values inferred for shocks around SN remnants \cite{ackermann2015b}.
For $\rs\sim$ 1 kpc and $\vs\sim10^{3}\,\kms$, $\emax\sim10^6$ GeV while for $\rs\sim50$ kpc and $\vs\sim500\,\kms$, $\emax$ reaches $10^8$ GeV.
This leads to the spectral break in the neutrino spectrum at $\enu\sim10^5$ GeV, as the production of $\enu$ is dominated by protons of energy $\ep\approx20\enu$ \cite{kelner2006}.
The observed photon spectrum cuts off at a much lower energy due to the attenuation of emitted $\gamma$-rays by electron-positron pair production on the cosmic UV-optical-infrared background photons out to the high redshifts $z>2$ where most quasars reside \cite{stecker2007}.
%
%
\\
\\
\indent
\textit{Multi-messenger implications.}---Assuming $pp$ interactions, the all flavor neutrino flux can be expressed in terms of the $\gamma$-ray flux, $\enu^2\Phi_{\nu}\approx6\eg^2\Phi_{\gamma}$ for $\enu\approx0.5\eg$ \cite{murase2013, zandanel2015, murase2016}.
This relation sets an upper limit on the power-law index of the accelerated protons \cite{murase2013}:
\begin{equation}
\gp\lesssim2+\frac{\ln\left[3\eg^2\Phi_{\gamma}|_{\eg}/(\enu^2\Phi_{\nu}|_{\enu})\right]}{\ln(2\enu/\eg)}\;.
\end{equation}
Given the most recent \textit{Fermi}-LAT data \cite{ackermann2015} and IceCube data \cite{aartsen2015}, we have verified that $\gp\lesssim 2.2-2.4$, in agreement with theoretical models \cite{caprioli2012, caprioli2014} and observations of SN remnant shocks \cite{ackermann2015b}.
If $\gp$ is taken beyond this limit, the EGB would be overproduced when attempting to accommodate the neutrino background.

Other astrophysical sources have been confirmed to produce neutrinos and may contribute to the CNB \cite{waxman2001, alvarez2002, mannheim2001, loeb2006, waxman1999, waxman1997}.
Blazars make up approximately half of the EGB at $\eg\lesssim10$ GeV and almost all the flux at higher photon energies.
They are estimated to explain the entire neutrino background at $\enu\gtrsim0.5$ PeV but only $\sim 10\%$ at lower energies, based on a leptohadronic model \cite{padovani2015}.
Star-forming galaxies produce $\sim 13\pm9\%$ of the EGB \cite{ackermann2012} via $pp$ interaction, indicating that they do not contribute significantly to the CNB for values of $\gp$ of interest \cite{tamborra2014}.
The central AGN in galaxy clusters is estimated to account for the neutrino background at $\enu\gtrsim0.1$ PeV, but not at lower energies \cite{fang2016}.
Additionally, the contribution from galaxy clusters to the EGB is only a few percent and thus negligible \cite{zandanel2015, fornasa2015}.
Other sources can be ruled out based on the $\gamma$-ray/neutrino branching ratio as they do not generate sufficient $\gamma$-ray emission to account for the EGB.
In comparison, the quasar outflow model can fully explain both the missing component of the EGB and the CNB.
The multi-messenger link between $\gamma$-ray and neutrino emission can be used to trace and confirm individual sources of neutrinos \cite{becker2005, murase2016}.
%
%
\\
\\
\indent
\textit{Summary.}---In this Letter, we adopted the quasar outflow parameters constrained by the best fit to the EGB data and calculated the simultaneous neutrino emission from these outflows.
The integrated neutrino flux of $\sim 10^{-7}\,\rm GeV\,s^{-1}\,cm^{-2}\,sr^{-1}$ at $\enu\approx$10 TeV, naturally explains the most recent IceCube data. 
The dominant mechanism for producing the $\gamma$-ray and neutrino emission is the interaction between protons accelerated by the outflow and the ambient interstellar protons.
In such a scenario, the branching ratio between $\gamma$-rays and neutrinos sets an upper limit on the power-law index of the accelerated proton distribution $\gp$ to be $\sim 2.2-2.4$ as inferred in SN remants \cite{ackermann2015b, caprioli2012, caprioli2014}.
Alternative sources such as blazars, star-forming galaxies and galaxy clusters can not account for both the $\gamma$-ray and neutrino backgrounds, while quasar outflows naturally explain both with a set of parameters consistent with direct observations of outflows \cite{tombesi2015} and SN remnants \cite{ackermann2015b}.
The inferred multi-messenger link can be used to constrain the high-energy physics of strong shocks at cosmological distances.
%

We thank Douglas Finkbeiner for helpful comments on the manuscript.
This work is supported by NSF grant AST-1312034.
%
%

%
\end{document}